\documentclass[a4paper,12pt]{article}
\usepackage{amsmath}
\usepackage{latexsym}
\usepackage{ascmac}
\usepackage{amsfonts}
\usepackage{amssymb}
\usepackage{latexsym}
\def\qed{\hfill $\Box$}
\usepackage{amsthm}

\theoremstyle{definition}
\newtheorem{thm}{Theorem}[section]
\newtheorem{defi}{Definition}[section]
\newtheorem{prop}{Proposition}[section]
\newtheorem{rem}{Remark}[section]

\newtheorem{cor}{Corollary}[section]
\newtheorem*{pf}{{\it Proof}}
\title{\bf A certain generalization of $q$-hypergeometric functions \\
and their related  monodromy preserving deformation}
\date{}
\author{K. Park\\ {\it Departmant of Mathematics, Graduate School of Science, Kobe University},\\{\it 1-1, Rokkodai, Nada-ku, Kobe} 657-8501, {\it Japan}\\Email: kpaku@math.kobe-u.ac.jp}
  \makeatletter
    \renewcommand{\theequation}{%
    \thesection.\arabic{equation}}
    \@addtoreset{equation}{section}
  \makeatother
\begin{document}
\maketitle
We define a series $\mathcal{F}_{M,N}$ as a certain generalization of $q$-hypergeometric function. We study its duality and the system of $q$-difference nonlinear equations which admits particular solutions in terms of $\mathcal{F}_{1,M}$.


 \section{Introduction}
  
    Tsuda \cite{tsuda10}, \cite{tsuda12} obtained a Hamiltonian system $\mathcal{H}_{N+1,M}$, which is an isomonodromy deformation of a certain $(N+1)\times (N+1)$ Fuchsian system on $\mathbb{P}^1$ with $M+3$ regular singularities, and which admits particular solutions in terms of a certain generalization of hypergeometric functions:
   \begin{equation}\label{gauss:kakutyou}
      F_{N+1,M} \Big({\{\alpha_j\},\{ \beta_i\} \atop \{ \gamma_j \}}; \{ x_i\}\Big)=\sum_{m_i \geq 0} \prod_{j=1}^N \frac{(\alpha_j)_{|m|}}{(\gamma_j)_{|m|}}\prod_{i=1}^M \frac{(\beta_i)_{m_i}}{(1)_{m_i}}\prod_{i=1}^M x_i^{m_i},
   \end{equation}
   where $1\leq i\leq M$, $1\leq j\leq N$, $(a)_n=\frac{\Gamma(a+n)}{\Gamma(a)}$ and, for a multi-index $m=(m_1,\cdots,m_M)$, $|m|=\sum_{i=1}^M m_i$.

      Our goal is to obtain a $q$-difference analogue  of the above result, and in this paper we give a partial result in this direction.  Namely, we will obtain a $q$-difference analogue $\mathcal{P}_{N,M}$ of Tsuda's system $\mathcal{H}_{N+1,M}$ in the case of $(M,N)=(M,1)$.

  The construction of the system $\mathcal{P}_{1,M}$ is as follows: Firstly, we define a series $\mathcal{F}_{M,N}$ as a generalization of the $q$-hypergeometric function ${}_2 \varphi_1$ \cite{Heine46} \cite{GasperRahman}. The series $\mathcal{F}_{M,N}$ has a duality relation (\ref{taishou}), which 
 gives an integral representation of $\mathcal{F}_{N,M}$.
  We derive a Pfaffian system from the integral representation of $\mathcal{F}_{1,M}$. Investigating the factorized structure of the coefficient matrices, we will interpret the Pfaffian system as a Lax pair 
  $T_z \overrightarrow{\Psi}=\overrightarrow{\Psi}A$, 
  $T_t \overrightarrow{\Psi}=\overrightarrow {\Psi}B$, where accessory parameters are written in terms of the function $\mathcal{F}_{M,N}$.
 Finally, as suggested by  this, we formulate  the system $\mathcal{P}_{1,M}$ which admits particular solutions of $\mathcal{F}_{1,M}$.
  %
  \begin{rem}  As mentioned in next section, $\mathcal{F}_{1,M}$ is equal to the $q$-Appell-Lauricella function $\varphi_D$. The system we will obtain in this paper can be identified with a system known as a $q$-difference analogue of the Garnier system \cite{sakai05}, \cite{sakai05-2}. 
  The $q$-Garnier system was first given  by Sakai \cite{sakai05}, and he presented particular solutions of the system in terms of $\varphi_D$ \cite{sakai05-2}. However, the present result is based on a different method  from that.
  \end{rem}
  This paper is organized as follows. 
  A definition of the series $\mathcal{F}_{M,N}$ and its fundamental properties are given in section $2$.
  A Pfaffian system which $\mathcal{F}_{1,M}$ satisfies is given in section $3$.
  We formulate the system $\mathcal{P}_{1,M}$   in section $4$. 
In the appendix we give a short summary of the result of Tsuda \cite{tsuda10}, \cite{tsuda12}.

  \section{The series  $\mathcal{F}_{N,M}$ as an extension of $q$-hypergeometric functions} 
  In this section, we define  the series $\mathcal{F}_{N,M}$ as an extension of $q$-hypergeometric functions and we show its duality relation and $q$-linear difference equations. The simplest case $\mathcal{F}_{1,1}$ corresponds to the Heine's function ${}_2 \varphi _1$ \cite{Heine46}, \cite{GasperRahman}.
    \begin{defi}We define a series $\mathcal{F}_{N,M}$ as
  \begin{gather} \label{Fmn}
\mathcal{F}_{N,M} \Big({\{a_j\},\{ b_i\} \atop \{ c_j \}}; \{ y_i\}\Big)=\displaystyle\sum_{m_i\geq 0}\prod_{j=1}^N \cfrac{(a _j)_{|m|} }{(c_j)_{|m|}}\prod _{i=1}^M \cfrac{(b_i)_{m_i}}{(q)_{m_i}}\prod_{i=1}^M y_i ^{m_i},
\end{gather}
  where
   $(a)_n=\frac{(a)_{\infty}}{(q^n a)_{\infty}}$. Here and in what follows the symbol $(a)_{\infty}$ means $(a)_{\infty}=\prod_{i=0}^{\infty} (1-q^i a)$. 
The series (\ref{Fmn}) converges in the region $|y_i|<1$ and  is continued analytically  to $|y_i|\geq1$.
  \end{defi}
 When $N=1$ or $M=1$, the series $(\ref{Fmn})$ is equal to the $q$-Appell-Lauricella function $\varphi_D$ or the generalized $q$-hypergeometric function ${}_{N+1} \varphi _N$, respectively:
\begin{equation}
\begin{array}{ll}
\mathcal{F}_{1,M} \displaystyle \left( {a,\{ b_i\} \atop c} ;\{ y_i \} \right)&=\displaystyle\sum_{m_i \geq 0} \frac{(a)_{|m|}}{(c)_{|m|}} \prod_{i=1}^M \frac{(b_i)_{m_i}}{(q)_{m_i}} y_i ^{m_i}
=\displaystyle\varphi_D \left( {a , \{b_i \} \atop c };\{ y_i \} \right),
\end{array}
\end{equation}
\begin{equation}
\begin{array}{ll}
\displaystyle \mathcal{F}_{N,1} \left( {\{a_j \} , b \atop \{c_j \} };y \right) &=\displaystyle\sum_{n \geq 0} \prod_{j=1}^N \frac{(a_j)_{m}(b)_{m}}{(c)_{m} (q)_{m}} y^{m} =\displaystyle{}_{N+1} \varphi_N \left( {\{a_j \} , b \atop \{ c_j \} }; y \right).
\end{array}
\end{equation}
There is a duality relation between the series $\mathcal{F}_{N,M}$ and $\mathcal{F}_{M,N}$ as follows:
\begin{prop}
The series $\mathcal{F}_{N,M}$ satisfies the relation
\begin{equation}\label{taishou}
\mathcal{F}_{N,M} \left( {\{ y_j \} ,\{a_i \} \atop \{b_jy_j \}};\{x_i \} \right) =\frac{(\{y_j\},\{a_i x_i \} )_{\infty}}{(\{b_jy_j \},\{ x_i \} )_{\infty}} \mathcal{F}_{M,N} \left( {\{x_i \} , \{b_j \} \atop \{a_i x_i \} };\{y_j \} \right).
\end{equation}
\end{prop}
\begin{pf}
We consider a double sum 
\begin{equation}\label{sum}
S=\sum_{m_i\geq 0} \sum_{n_j\geq 0} \prod_{i=1}^M \frac{(a_i)_{m_i}}{(q)_{m_i}} x_i ^{m_i} \prod_{j=1}^N \frac{(b_j)_{n_j}}{(q)_{n_j}} y_j ^{n_j} q^{|m||n|}.
\end{equation}
Using the $q$-binomial theorem
\begin{equation}
\sum_{n\geq 0} \frac{(b)_n}{(q)_n}z^n=\frac{(bz)_{\infty}}{(z)_{\infty}},
\end{equation}
we can take the sum  over $n_j\geq 0$ in (\ref{sum}). Then the sum $S$ is represented by the series $\mathcal{F}_{N,M}$:
\begin{align}
S&=\sum_{m_i\geq 0} \prod_{i=1}^M \frac{(a_i)_{m_i}}{(q)_{m_i}} x_i^{m_i} \prod_{j=1}^N \frac{(b_jy_j)_{\infty}}{(y_j)_{\infty}} \nonumber\\
 &=\prod_{j=1}^N \frac{(b_j y_j)_{\infty}}{(y_j)_{\infty}}
 \sum_{m_i\geq 0}\prod_{i=1}^M \frac{(a_i)_{m_i}}{(q)_{m_i}} x_i^{m_i} \prod_{j=1}^N \frac{(y_j)_{|m|}}{(b_jy_j)_{|m|}}\nonumber\\
 &=\prod_{j=1}^N \frac{(b_jy_j)_{\infty}}{(y_j)_{\infty}} \mathcal{F}_{N,M} \left( {\{ y_j \},\{a_i\} \atop \{b_j y_j \} };\{x_i\} \right)\label{nwa}.
 \end{align}
 Similarly, when we sum over $m_i \geq 0$, the sum $S$ is represented by the series $\mathcal{F}_{M,N}$
 \begin{equation}\label{mwa}
S=\prod_{i=1}^M \frac{(a_ix_i)_{\infty}}{(x_i)_{\infty}} \mathcal{F}_{M,N} \left( {\{x_i \},\{b_j \} \atop \{ a_ix_i\} };\{ y_j \} \right).
 \end{equation}
 From (\ref{nwa}), (\ref{mwa}), we obtain the relation $(\ref{taishou})$. 
\qed
\end{pf}
\begin{rem}
When $N=1$ or $M=1$, the relation (\ref{taishou}) is known in \cite{And72} and \cite{KN03}.
\end{rem}
We can interpret the equation (\ref{taishou}) as
an integral representation of $\mathcal{F}_{N,M}$ as follows.
  \begin{cor}\label{cor:sekibun}
  With $y_j=q^{\gamma_j}$, the relation $(\ref{taishou})$ can be rewritten as
  \begin{equation}\label{sekibun}
  \mathcal{F}_{N,M} \left( {\{ q^{\gamma_j} \} ,\{a_i \} \atop \{b_j q^{\gamma_j} \}};\{x_i \} \right)=\frac{(\{q^{\gamma_j}\},\{b_j\})_{\infty}}{(\{b_j q^{\gamma_j} \},q)_{\infty}} \prod_{j=1}^N \int _0^1d_q t_j \prod_{i=1}^M \frac{(a_ix_i \prod_{j=1}^N t_j )_{\infty}}{(x_i \prod_{j=1}^N t_j)_{\infty}} \prod_{j=1}^N \frac{(qt_j)_{\infty}}{(b_jt_j)_{\infty}} \frac{t_j ^{\gamma_j-1}}{1-q},
  \end{equation}
  where the Jackson integral is defined as
  \begin{equation}
  \int_{0}^c d_q t f(t) = c(1-q) \sum_{n\geq 0} f(cq^n) q^n.
  \end{equation}
  \end{cor}
  \begin{pf}
  The right-hand side of (\ref{taishou})  can be written as
  \begin{align}
  (RHS)=&\displaystyle\frac{(\{ q^{\gamma_j}\},\{a_ix_i \} )_{\infty}}{(\{ b_j q^{\gamma_j}\},\{x_i\} )_{\infty}}\sum_{n_j\geq 0} \prod_{i=1}^M \frac{(x_i)_{|n|}}{(a_ix_i)_{|n|}}\prod_{j=1}^N \frac{(b_j)_{n_j}}{(q)_{n_j}} (q^{\gamma_j})^{n_j}\nonumber\\
  =&\displaystyle \frac{(\{q^{\gamma_j}\},\{b_j\})_{\infty}}{(\{b_j q^{\gamma_j} \},q)_{\infty}}
  \sum_{n_j\geq 0} \prod_{i=1}^M \frac{(a_ix_iq^{|n|})_{\infty}}{(x_i q^{|n|})_{\infty}} \prod_{j=1}^N \frac{(q^{n_j+1})_{\infty}}{(b_jq^{n_j})_{\infty}} q^{n_j \gamma_j}\nonumber\\
  =&\frac{(\{q^{\gamma_j}\},\{b_j\})_{\infty}}{(\{b_j q^{\gamma_j} \},q)_{\infty}} \prod_{j=1}^N \int _0^1d_q t_j \prod_{i=1}^M \frac{(a_ix_i \prod_{j=1}^N t_j )_{\infty}}{(x_i \prod_{j=1}^N t_j)_{\infty}} \prod_{j=1}^N \frac{(qt_j)_{\infty}}{(b_jt_j)_{\infty}} \frac{t_j ^{\gamma_j-1}}{1-q}.
  \end{align}
  Hence, we obtain 
  $(\ref{sekibun})$. 
  \qed
  \end{pf}
    \begin{prop}
  The series $\mathcal{F}=\mathcal{F}_{N,M} \Big({\{a_j\},\{ b_i\} \atop \{ c_j \}}; \{ y_i\}\Big)$ satisfies the  $q$-difference equations
  \begin{align}
  &\left\{ \prod_{j=1}^N (1-c_jq^{-1} T) \cdot (1-T_{y_s}) y_s-  \prod_{j=1}^N (1-a_jT)\cdot (1-b_s T_{y_s})\right\}\mathcal{F}=0\quad(1\leq s \leq M),\label{sabun1}\\
  &\{y_s (1-b_r T_{y_s})(1-T_{y_s}) -y_s (1-b_s T_{y_s})(1-T_{y_r}) \} \mathcal{F}=0\quad (1\leq r <s\leq M),\label{sabun2}
  \end{align}
  where $T_{y_s}$ is the $q$-shift operator for  the variable $y_s$ and $T=T_{y_1}\cdots T_{y_M}$.
  \end{prop}
  \begin{pf}
  The coefficient of $\prod_{i=1}^M y_i^{m_i}$: $C(\{ m_i \})$ satisfies the following relation
  \begin{equation}\label{keisu}
  \prod_{j=1}^N(1-q^{|m|}c_j)\cdot(1-q^{m_s+1})C(\{m_i + \delta_{i,s}\}) =\prod_{j=1}^N (1-q^{|m|}a_j)\cdot(1-q^{n_s}b_s)C(\{ m_i\}).
  \end{equation}
  Multiplying both sides of $(\ref{keisu})$ by $\prod_{i=1}^M y_i^{m_i}$ and summing them over $m_i\geq 0$, we obtain $(\ref{sabun1})$.
The equation (\ref{sabun2}) is obtained similarly.
    \qed
  \end{pf}
  \section{A Pfaffian system derived from $\mathcal{F}_{1,M}$}
  In this section, we derive a Pfaffian system from an integral representation of $\mathcal{F}_{1,M}$, and then
  we derive a $2\times 2$ system by a certain specialization.
  \subsection{A Pfaffian system}
  We derive a Pfaffian system which the integral represantation of  $\mathcal{F}_{1,M}$ satisfies. In this subsection, we basically follow the method given in \cite{mimachi89}. From Cor. \ref{cor:sekibun}, the integral representation of $\mathcal{F}_{1,M}$ is given as
  \begin{equation}\label{sekibun1}
  \mathcal{F}_{1,M} \left( { q^{\gamma_1}  ,\{a_i \} \atop b_1q^{\gamma_1} };\{x_i \} \right)=\frac{(q^{\gamma_1},b_1)_{\infty}}{(b_1 q^{\gamma_1} ,q)_{\infty}}  \int _0^1d_q u_1 \prod_{i=1}^M \frac{(a_ix_i  u_1 )_{\infty}}{(x_i  u_1)_{\infty}} \frac{(qu_1)_{\infty}}{(b_1u_1)_{\infty}} \frac{u_1 ^{\gamma_1}}{1-q}.
  \end{equation}
  We set the integrand in $(\ref{sekibun1})$ as
  \begin{equation}\label{integrand}
  \Phi(u_1)=u_1^{\gamma_1} \prod_{i=0}^M \frac{(a_ix_iu_1)_{\infty}}{(x_iu_1)_{\infty}}\quad(x_0=b_1, a_0x_0=q).
  \end{equation}
  Following \cite{mimachi89}, 
  we define functions $\Psi_0$, $\Psi_{1,i}$ $(1\leq i \leq M)$ as 
  \begin{align}
  \Psi_0&=\langle \Phi p_0\rangle,\\
  \Psi_{1,i}&=\langle \Phi p_{1,i} \rangle \quad (1\leq i\leq M),
  \end{align}
  where $p_0=1$, $p_{1,i}=\frac{1-x_0u_1}{1-a_ix_iu_1}\prod_{k=1}^{i-1}\frac{1-x_ku_1}{1-a_kx_ku_1}$ and $\langle \  \rangle$ means a kind of  Jackson integral for $u_1$. 
  Namely, $\langle f(u_1) \rangle=\displaystyle\sum_{n \in \mathbb{Z}} f(q^n)$.
  We will see  that $\{ \Psi_0, \Psi_{1,1},\cdots, \Psi_{1,M} \}$ is a basis of a solution of the Pfaffian system. 
  
  We define an exchange operator $\sigma_i$ acting on a function $f$ of $\{x_i, a_i \}$ as
  \begin{equation}
  \sigma_i (f) =f|_{x_i\leftrightarrow x_{i+1}, a_i\leftrightarrow a_{i+1}}.
  \end{equation}
  We note that $\sigma_i (\Phi (u_1))=\Phi (u_1)$. When the operator $\sigma_i$ acts on functions $p_0$, $p_{1,i}$ $(1\leq i \leq M)$, we have the following relations.
  \begin{prop}\label{prop:gokan} We have
  \begin{gather}
  \sigma_i 
  \begin{bmatrix}
  p_{1,i}\\
  p_{1,i+1}
  \end{bmatrix}
  =\frac{1}{x_i-a_{i+1}x_{i+1}} 
  \begin{bmatrix}
  (1-a_i)x_i& a_ix_i-a_{i+1}x_{i+1}\\
  x_i-x_{i+1}&(1-a_{i+1})x_{i+1}
  \end{bmatrix}
  \begin{bmatrix}
  p_{1,i}\\
  p_{1,i+1}
  \end{bmatrix},\label{gokan1}\\
  \sigma_i(p_{1,k})=p_{1,k}\ (k \neq i, i+1).\label{gokan2}
  \end{gather}
  \end{prop}
  \begin{pf}
  Consider the  equations
  \begin{align}
  \sigma_i (p_{1,i}) &=s_{11} \ p_{1,i}+s_{12}\ p_{1,i+1},\\
  \sigma_i (p_{1,i+1})&= s_{21}\ p_{1,i}+s_{22}\ p_{1,i+1}.
  \end{align}
  One can show that there exist unique coefficients $s_{11}$, $s_{12}$, $s_{21}$, $s_{22}$ independent of $u_1$ satisfying these relations. Hence
  we obtain the form (\ref{gokan1}). The equation (\ref{gokan2}) is obvious.
  \qed
  \end{pf}
  For the action of the $q$-shift operator $T_{x_M}$ on $\Psi_0$, $\cdots$, $\Psi_{1,M}$, we obtain the following equations.
  \begin{prop}\label{prop:mup}We have
  \begin{align}
  &T_{x_M} (\Psi_0)=\frac{x_M-x_0}{a_Mx_M-x_0} \rho (\Psi_0)+\frac{(a_M-1)x_M}{a_Mx_M-x_0}\rho(\Psi_{1,1}),\label{shift1}\\
  &T_{x_M} (\Psi_{1,i})=\rho(\Psi_{1,i+1})\quad(i\neq M),\label{shift2}\\
  &T_{x_M} (\Psi_{1,M})=\frac{q^{-\gamma_1}(1-x_0)}{a_Mx_M-x_0} \left[ \rho (\Psi_0)+\frac{a_Mx_M-1}{1-x_0} \rho (\Psi_{1,1})\right],\label{shift3}
  \end{align}
  where $\rho=\sigma_{M-1}\cdots\sigma_1$.
  \end{prop}
  \begin{pf}
  First, we make a shift by  $T_{x_M}$ on $\Phi p_0$, $\Phi p_{1,i}$ $(1\leq i \leq M)$. We easily obtain the following equations
  \begin{align}
    &T_{x_M} (\Phi p_{1,i})=\Phi \rho (p_{1,i+1}),\label{1i}\\
  &T_{x_M} (\Phi p_0) =\Phi \frac{1-x_Mu_1}{1-a_Mx_Mu_1}\label{p0},\\
  &T_{x_M} T_{u_1}^{-1} (\Phi p_{1,M}) =q^{-\gamma_1} \Phi \frac{1-u_1}{1-a_Mx_Mu_1}\label{p1M}.
  \end{align}
  The right-hand side of the equations $(\ref{p0})$, $(\ref{p1M})$ can be rewritten as a linear combination of $\rho(\Phi p_0)$ and $\rho (\Phi p_{1,1})$ respectively, that is,
  \begin{align}
  &T_{x_M}(\Phi p_0)=\frac{x_M-x_0}{a_Mx_M-x_0} \rho (\Phi p_0) + \frac{(a_M-1)x_M}{a_Mx_M-x_0}\rho (\Phi p_{1,1}),\label{0}\\
  &T_{x_M} T_{u_1}^{-1} (\Phi p_{1,M}) =q^{-\gamma_1} \left(\frac{1-x_0}{a_Mx_M-x_0} \rho (\Phi p_0)+\frac{a_Mx_M-1}{a_M x_M -x_0}\rho (\Phi p_{1,1}) \right)\label{M}.
  \end{align}
  Integrating the equations $(\ref{1i})$, $(\ref{0})$ and $(\ref{M})$ with respect to  $u_1$, we obtain the equations $(\ref{shift1})$-$(\ref{shift3})$. 
  \qed
  \end{pf}
   Combining Prop. \ref{prop:gokan} and Prop. \ref{prop:mup}, we obtain the following theorem.
  \begin{thm}\label{thm:pfaff}
  The vector $\overrightarrow{\Psi}=\begin{bmatrix}\Psi_0,\Psi_{1,1},\cdots,\Psi_{1,M}\end{bmatrix}$ satisfies the  equation
  \begin{gather}\label{appu}
  T_{x_M} \overrightarrow{\Psi}=\overrightarrow{\Psi} A_M=\overrightarrow{\Psi} R_{M-1}R_{M-2}\cdots R_1 Q,
  \end{gather}
  where the matrices $R_i$ and $Q$ are given as
\begin{gather}
R_i=
\begin{bmatrix}E_{i}&&&\\&\frac{(1-a_i)x_i}{x_i-a_Mx_M}&\frac{x_i-x_{M}}{x_i-a_{M}x_{M}}&\\&&&E_{M-i-1}\end{bmatrix},\\
Q=
\begin{bmatrix}\frac{x_M-x_0}{a_Mx_M-x_0}&&\frac{q^{-\gamma_1}(1-x_0)}{a_Mx_M-x_0}\\ \frac{(a_M-1)x_0}{a_Mx_M-x_0}&&\frac{q^{-\gamma_1}(a_Mx_M-1)}{a_Mx_M-x_0}\\&E_{M-1}&\end{bmatrix}.
\end{gather}
$E_i$ is the unit matrix of dimension $i$.
\end{thm}
 \begin{rem}
 The equations for the other variables 
 \begin{equation}\label{appui}
  T_{x_i}  \overrightarrow{\Psi}=\overrightarrow{\Psi}A_i,
  \end{equation}
  for $i=1,\cdots, M-1$
 can be derived from (\ref{appu}) and the action of $\{ \sigma_i\}$ in Prop. \ref{prop:gokan}.
 The coefficient matrices in (\ref{appu}) and (\ref{appui}) 
satisfy a compatibility condition
\begin{equation}
A_i (T_{x_i} A_j)=A_j(T_{x_j}A_i).
\end{equation}
\end{rem} 
\begin{rem}
When $M=1$, the equation is   just solved by the ${}_2 \varphi_1$ function.
In fact we obtain,  by Theorem \ref{thm:pfaff}, the equation 
\begin{equation}\label{nm11-2}
T_{x_1} \begin{bmatrix}\Psi_0, \Psi_{1,1}\end{bmatrix}= \begin{bmatrix}\Psi_0, \Psi_{1,1}\end{bmatrix} \begin{bmatrix}\frac{x_1-x_0}{a_1x_1-x_0}&\frac{q^{-\gamma_1}(1-x_0)}{a_1x_1-x_0}\\ \frac{(a_1-1)x_0}{a_1x_1-x_0}&\frac{q^{-\gamma_1}(a_1x_1-1)}{a_1x_1-x_0}\end{bmatrix}.
\end{equation}
\end{rem}
 \subsection{A reduction to $2\times 2$ form}
 In this section we reduce the equation (\ref{appui}) into $2\times2$ form. To do this, we specialize the parameter $a_M$ to be $1$. Then the integrand (\ref{integrand}), and  hence $\Psi_{1,i}$ $(1\leq i \leq M-1)$ and $\Psi_0$, become independent of $x_M$.
  Therefore we can consider $\Psi_{1,i}$ $(1\leq i \leq M-1)$ as $\Psi_0$ times $r_i$, where $r_i$ is a rational function in $x_0$, $\cdots$, $x_{M-1}$. Based on this fact, we have the following.
  \begin{thm}\label{thm:Lax}
Specializing $a_M=1$ and setting $z=x_M$, $t=x_{M-1}$ and $\Psi_{1,i}=r_i \Psi_0\quad(1\leq i \leq M-1)$,
the equations in Theorem \ref{thm:pfaff} can be rewritten as 
  \begin{equation}
  \begin{cases}\label{2kake2}
  T_{z} (\Psi_0,\Psi_{1,M})=(\Psi_0,\Psi_{1,M}) \left(\displaystyle\prod_{i=1}^{M-1}
  \begin{bmatrix}1/r_i&1\\ z&r_ix_i\end{bmatrix}^{-1}
  \begin{bmatrix}1/r_i&1\\z&a_ir_ix_i\end{bmatrix} \right)
  \begin{bmatrix}1&1\\z&x_0\end{bmatrix}^{-1}
  \begin{bmatrix}1&1\\z&1\end{bmatrix}
  \begin{bmatrix}1&0\\0&q^{-\gamma_1} \end{bmatrix},\\
  T_{t} (\Psi_0,\Psi_{1,M})=(\Psi_0,\Psi_{1,M}) \begin{bmatrix}u&1\\z/q&a_{M-1}t/u\end{bmatrix}^{-1}\begin{bmatrix}u&1\\z/q&t/u\end{bmatrix},
  \end{cases}
  \end{equation}
  where $u$ is independent of $z$.
  \end{thm}
  \begin{pf}When $a_M=1$, obviously we have $T_z (\Psi_0)=\Psi_0$. We will compute $T_z(\Psi_{1,M})$.
In  equation (\ref{appu}), 
the coefficient matrices become
 \begin{gather}
 R_i=\begin{bmatrix}E_i&&&\\&\frac{(1-a_i)x_i}{x_i-z}&1&\\&\frac{a_ix_i-z}{x_i-z}&0&\\&&&E_{M-i-1}\end{bmatrix},\quad Q=\begin{bmatrix}1&&\frac{q^{-\gamma_1}(1-x_0)}{z-x_0}\\0&&\frac{q^{-\gamma_1}(z-1)}{z-x_0}\\&E_{M-1}&\end{bmatrix}.
 \end{gather}
We rewrite the equation $(\ref{appu})$:
\begin{equation}\label{vector}
\begin{array}{rl}
\overrightarrow{\Psi}^{(1)}:=&\overrightarrow{\Psi} R_{M-1}=(\Psi_0,\cdots,\Psi_{1,M-2},\Psi_{1,M-1}^{(1)},\Psi_{1,M-1}), \\
&\Psi_{1,M-1}^{(1)}:=a(M-1)\Psi_{1,M-1}+b(M-1)\Psi_{1,M},\\[3mm]
\overrightarrow{\Psi}^{(2)}:=&\overrightarrow{\Psi}^{(1)} R_{M-2}=(\Psi_0,\cdots,\Psi_{1,M-2}^{(2)},\Psi_{1,M-2},\Psi_{1,M-1}), \\
&\Psi_{1,M-2}^{(2)}:=a(M-2)\Psi_{1,M-2}+b(M-2)\Psi_{1,M-1}^{(1)},\\
\vdots \\
\overrightarrow{\Psi}^{(M-1)}:=&\overrightarrow{\Psi}^{(M-2)} R_{1}=(\Psi_0,\Psi_{1,1}^{(M-1)},\Psi_{1,1}\cdots,\Psi_{1,M-1}), \\
&\Psi_{1,1}^{(M-1)}:=a(1)\Psi_{1,1}+b(1)\Psi_{1,2}^{(M-2)},\\[3mm]
\overrightarrow{\Psi}^{(M)}:=&\overrightarrow{\Psi}^{(M-1)} Q=(\Psi_0,\Psi_{1,1},\cdots,\Psi_{1,M-1},\Psi_{1,M-1}^{(M)}), \\
&\Psi_{1,M-1}^{(M)}:=a(M)\Psi_{1,1}+b(M)\Psi_{1,1}^{(M-1)},
\end{array}
\end{equation}
where the coefficients $a(i)$ and $b(i)$ are as 
\begin{equation}\label{aibi}
a(i)=
\begin{cases}
\frac{(1-a_i)x_i}{x_i-z}\quad (1\leq i\leq M-1),\\
\frac{q^{-\gamma_1}(1-x_0)}{z-x_0}\quad (i=M),
\end{cases}
b(i)=
\begin{cases}
\frac{a_ix_i-z}{x_i-z}\quad (1\leq i\leq M-1),\\
\frac{q^{-\gamma_1}(1-z)}{z-x_0}\quad (i=M).
\end{cases}
\end{equation}
We obtain the following equation  from (\ref{vector}) and (\ref{aibi}):
\begin{equation}\label{rank2}
T_z (\Psi_{1,M})=\Psi_{1,M-1}^{(M)}.
\end{equation}
The  equation $(\ref{rank2})$ can be rewritten as
\begin{equation}
\begin{array}{rl}
\Psi_{1,M-1}^{(M)}=&\displaystyle a(M)\Psi_0+b(M) \sum_{i=1}^{M-1} a(i)\prod_{j=1}^{i-1} b(j)  \Psi_{1,i}+b(M) \prod_{k=1}^{M-1} b(k)\Psi_{1,M}\\
=&\displaystyle \biggl(a(M)+b(M)\sum_{i=1}^{M-1} r_i a(i)\prod_{j=1}^{i-1} b(j)\biggr) \Psi_0+b(M) \prod_{k=1}^{M-1} b(k)\Psi_{1,M}.
\end{array}
\end{equation}
From the above, we obtain the first equation of equation $(\ref{2kake2})$:
\begin{equation}
\begin{array}{l}
T_{z} (\Psi_0,\Psi_{1,M})\\
=(\Psi_0,\Psi_{1,M}) \begin{bmatrix}1&\displaystyle a(M)+b(M)\biggl( \sum_{i=1}^{M-1} r_i a(i) \prod_{j=1}^{i-1} b(j) \biggr)\\0&\displaystyle b(M)\biggl( \prod_{k=1}^{M-1} b(k)\biggr)\end{bmatrix}\\[15mm]
  =(\Psi_0,\Psi_{1,M})\left( \displaystyle \prod_{i=1}^{M-1}\begin{bmatrix}1&r_{M-i}a(M-i)\\0&b(M-i)\end{bmatrix} \right)
  \begin{bmatrix}1&a(M)\\0&b(M)\end{bmatrix}\\[10mm]
  =(\Psi_0,\Psi_{1,M})\left(\displaystyle\prod_{i=1}^{M-1} \begin{bmatrix}1/r_i&1\\ z&r_ix_i\end{bmatrix}^{-1}
  \begin{bmatrix}1/r_i&1\\z&a_ir_ix_i\end{bmatrix}\right)
  \begin{bmatrix}
  1&1\\
  z&x_0
  \end{bmatrix}^{-1}
  \begin{bmatrix}
  1&1\\
  z&1
  \end{bmatrix}
  \begin{bmatrix}
  1&0\\
  0&q^{-\gamma_1}
  \end{bmatrix}.
  \end{array}{}
  \end{equation}
  The second equation $T_{t} (\Psi_0,\Psi_{1,M})$ is obtained similarly by considering $T_{t}=\sigma_{M-1} T_z \sigma_{M-1}$.
  \qed
  \end{pf}
  \section{A monodromy preserving deformation $\mathcal{P}_{1,M}$ related to $\mathcal{F}_{1,M}$}
  In this section, we interpret  the equations (\ref{2kake2}) as a  Lax pair for nonlinear 
  equations associated with special solutions. As suggested by this, we obtain the system $\mathcal{P}_{1,M}$ we are aiming at.
\begin{defi}\label{system}
We consider the following Lax form for the unknown function $\Psi(z)=[\Psi_1(z),\Psi_2(z)]$
\begin{equation}\label{Laxpair}
\begin{cases}
{\Psi(qz)}={\Psi} A,\\
\overline{\Psi}={\Psi}B,
\end{cases}
\end{equation}
where the coefficient matrices are
 \begin{equation}\begin{cases}
 A=d X_1^{-1} X_2\cdots X_{2M-1}^{-1} X_{2M},\label{matrixA}\\
 B=X_{2M}(z/q)^{-1} X_{2M-1} (z/q),
 \end{cases}\end{equation}
  and $d$ is a diagonal matrix and  
 $X_i=\begin{bmatrix}
 u_i&1\\
 z&c_i/u_i
 \end{bmatrix}$. We denote by $\overline{*}$  a "time-evolution" defined as $\{ c_i \to q c_i, (i=2M-1,2M)\}$.
 Then we define a system of nonlinear $q$-difference equations $\mathcal{P}_{1,M}$  for the unknown variables $u_i$ $(1\leq i \leq 2M)$, through the compatibility conditions
 \begin{equation}\label{ryoritsu2}
 A(z) B(qz)=B(z) \overline{A(z)}.
 \end{equation}
\end{defi}
\begin{prop}The equation (\ref{ryoritsu2}) is satisfied if and only if 
the variables $\{ u_i\}$ and  $\{ \overline{u_i} \}$ 
are related by a certain  birational mapping.
\end{prop}
\begin{pf} First, we note that, if there exist $\{ \overline{u_i}\}$ which satisfy (\ref{ryoritsu2})  for given $\{ u_i\}$, then such $\{ \overline{u_i}\}$ are unique. This uniqueness follows  by considering the kernel of both sides at $z=c_i$.
%
The existence of the birational mapping $\{ u_i \} \mapsto \{ \overline{u_i}\}$
is obtained as follows. Consider the following simple transformations:
\begin{equation}
\begin{array}{c}
u_k\mapsto u_k'=\displaystyle\frac{-c_lu_k+c_ku_l}{(u_k-u_l)u_k}, u_l\mapsto u_l'=\frac{-c_lu_k+c_ku_l}{(u_k-u_l)u_k},\\
\text{such that} \begin{bmatrix}u_k&1\\z&c_k/u_k\end{bmatrix} \begin{bmatrix}u_l&1\\z&c_l/u_l\end{bmatrix}^{-1}=\begin{bmatrix}u_l'&1\\z&c_l/u_l'\end{bmatrix}^{-1}\begin{bmatrix}u_k'&1\\z&c_k/u_k'\end{bmatrix}.
\end{array}
\end{equation}
\begin{equation}
\begin{array}{c}
u_k\mapsto u_k'=\displaystyle\frac{u_l(c_k+u_ku_l)}{c_l+u_ku_l}, u_l\mapsto u_l'=\frac{u_k(c_l+u_ku_l)}{c_k+u_ku_l},\\
\text{such that} \begin{bmatrix}u_k&1\\z&c_k/u_k\end{bmatrix} \begin{bmatrix}u_l&1\\z&c_l/u_l\end{bmatrix}=\begin{bmatrix}u_l'&1\\z&c_l/u_l'\end{bmatrix}\begin{bmatrix}u_k'&1\\z&c_k/u_k'\end{bmatrix},\\
\end{array}
\end{equation}
and
\begin{equation}
\begin{array}{c}
u \mapsto u'=\displaystyle\frac{d_1 u}{d_2}, v\mapsto v'=\displaystyle\frac{d_1 v}{d_2},\\
\text{such that} \begin{bmatrix}u&1\\z/q&c/u\end{bmatrix}^{-1}\begin{bmatrix}v&1\\z/q&d/v \end{bmatrix}\begin{bmatrix}d_1&0\\0&d_2\end{bmatrix} 
=\begin{bmatrix}d_1&0\\0&d_2\end{bmatrix} \begin{bmatrix}u'&1\\z&qc/u' \end{bmatrix}^{-1}\begin{bmatrix}v'&1\\z&qd/v'\end{bmatrix}.\\
\end{array}
\end{equation}
By composing them we obtain a birational mapping $\{ u_i\} \mapsto \{ \overline{u_i} \}$ which satisfies (\ref{ryoritsu2}). \qed
\end{pf}
\begin{rem}
The system $\mathcal{P}_{1,M}$ obtained from the equation $(\ref{ryoritsu2})$ is equivalent to the $q$-Garnier system \cite{sakai05}, \cite{NY17}.
\end{rem}
 \begin{prop}\label{thm:tokushuka}The coefficient matrices of (\ref{2kake2}) have the forms in (\ref{matrixA}).
 \end{prop}
 \begin{pf}
We will rewrite the coefficient matrix of the first equation of (\ref{2kake2}). 
We consider the relevant five factors of it. Then we obtain
\begin{equation}
\begin{array}{l}
  \begin{bmatrix}1/r_{M-1}&1\\ z&r_{M-1}t\end{bmatrix}^{-1}
  \begin{bmatrix}1/r_{M-1}&1\\z&a_{M-1}r_{M-1}t\end{bmatrix}
  \begin{bmatrix}1&1\\z&x_0\end{bmatrix}^{-1}
  \begin{bmatrix}1&1\\z&1\end{bmatrix}
  \begin{bmatrix}1&0\\0&q^{-\gamma_1} \end{bmatrix}\\
=
  \begin{bmatrix}p&1\\z&x_0/p\end{bmatrix}^{-1}
  \begin{bmatrix}p&1\\z&1/p\end{bmatrix}
  \begin{bmatrix}1&0\\0&q^{-\gamma_1} \end{bmatrix}
  \begin{bmatrix}u&1\\ z&t/u\end{bmatrix}^{-1}
  \begin{bmatrix}u&1\\z&a_{M-1}t/u\end{bmatrix}\\
=
  \begin{bmatrix}p&1\\z&x_0/p\end{bmatrix}^{-1}
  \begin{bmatrix}p&1\\z&1/p\end{bmatrix}
  \begin{bmatrix}1&0\\0&q^{-\gamma_1} \end{bmatrix}\cdot
  B(qz,t)^{-1},
\end{array}
\end{equation}
where $p=q^{-\gamma_1}\frac{(1-a_{M-1}t)r_{M-1}+x_0-1}{(-1+(1-a_{M-1})r_{M-1})t+x_0}u$.
 From this we see that the coefficient matrices of (\ref{2kake2}) take  the form in (\ref{matrixA}).
 \qed
 \end{pf}
 As a corollary of  Prop. \ref{thm:tokushuka} we have the following theorem.
 \begin{thm}
 The system $\mathcal{P}_{1,M}$ admits a particular solution in terms of $\mathcal{F}_{1,M}$. 
 \end{thm}
 \begin{rem}
For the case of $M=2$, the equation (\ref{ryoritsu2}) is equivalent to $q$-Painlev{\'e} VI \cite{sakai96}. 
In fact, the Lax pair (\ref{Laxpair}), (\ref{matrixA}) is 
\begin{equation}
\begin{array}{l}
A=\begin{bmatrix}u_1&1\\z&c_1/u_1\end{bmatrix}^{-1}\begin{bmatrix}u_2&1\\z&c_2/u_2\end{bmatrix}\begin{bmatrix}u_3&1\\z&c_3/u_3\end{bmatrix}^{-1}\begin{bmatrix}u_4&1\\z&c_4/u_4\end{bmatrix},\\
B=\begin{bmatrix}u_4&1\\z/q&c_4/u_4\end{bmatrix}^{-1}\begin{bmatrix}u_3&1\\z/q&c_3/u_3\end{bmatrix}.
\end{array}
\end{equation}
Solving the  compatibility condition  
\begin{equation}
A(z) B (qz)=B(z) \overline{A(z)},
\end{equation}
we obtain the following $q$-$P_{VI}$ equation for  $f$, $g$:
\begin{equation}
\begin{array}{l}
\overline{f}f=\displaystyle-\frac{c_3(c_2-g)(a^2 c_4 g-c_1c_3)q}{(ag-c_3) (ag-c_3q)},\\
\overline{g}g=\displaystyle\frac{c_3^2(c_1-\overline{f})(c_2-\overline{f})q^2}{a^2(c_3q-\overline{f})(c_4q-\overline{f})},
\end{array}
\end{equation}
where
\begin{equation}
f=\frac{c_1c_4u_2^2u_3+c_2c_4u_1u_3^2-c_1c_4u_2u_3^2-c_1c_3u_2^2u_4}{c_4u_1u_2u_3-c_3u_1u_2u_4+c_2u_1u_3u_4-c_1u_2u_3u_4} ,\quad g=\frac{c_3 (c_2 u_1-c_1 u_2)u_4^2}{u_1^2 (c_3u_4-c_4u_3)},
\end{equation}
%
and $a=\frac{u_1u_3}{u_2u_4}$ is a constant.
\end{rem}
%
\section*{Acknowledgement} The author would like to express her gratitude to Professor Yasuhiko Yamada for valuable suggestions and encouragement. She is also grateful to Professor Wayne Rossman for careful reading the manuscript and several improvements.

\renewcommand{\theequation}{A.\arabic{equation} }
  \setcounter{equation}{0}

  \section*{Appendix} In this appendix we give a short summary of the result of Tsuda \cite{tsuda10}, \cite{tsuda12}. 
   Tsuda derived a Hamiltonian system $\mathcal{H}_{N+1,M}$ which admits particular solutions in terms of a certain generalization of hypergeometric functions from the following Lax formalism \cite{tsuda10}.
   Consider an $(N+1)\times (N+1)$ Fuchsian system
   \begin{equation}\label{TsudaA}
   \frac{\partial \Phi}{\partial z} =A \Phi =\sum _{i=0}^{M+1} \frac{A_i}{z-u_i} \Phi,
   \end{equation}
   with $M+3$ regular singularities $\{ u_0=1,u_1,\cdots, u_M, u_{M+1}=0,u_{M+2}=\infty \} \subset \mathbb{P}^{1}$, of which the characteristic exponents at each singularity $z=u_i$ 
   are listed in the following table (Riemann scheme):
   
   \begin{table}[h]
\centering
\begin{tabular}{cc}
\hline
Singularity          & Exponents                                       \\ \hline
$u_i (0\leq i \leq M)$ & $(-\theta_i,0,\cdots,0)$                          \\
$u_{M+1}=0$            & $(e_0, e_1,\cdots, e_{N})$                        \\
$u_{M+2}=\infty$       & $(\kappa_0-e_0,\kappa_1-e_1,\cdots,\kappa_N-e_N)$ \\ \hline
\end{tabular}
\end{table}
We have Fuchs' relation $\sum _{n=0}^N \kappa_n=\sum _{i=0}^M \theta _i $.
We can normalize the exponents as $\sum _{n=0}^{N} e_n=\frac{N}{2}$ without loss of generality.  
Such Fuchsian systems as above then turn out to constitute a $2MN$-dimensional family and can be written in terms of the accessory parameters $b_n^{(i)}$ and $c_n^{(i)}$ in the following way:
\begin{equation}
\begin{array}{l}
A_i={}^T (b_0^{(i)},b_1^{(i)},\cdots,b_N^{(i)})\cdot (c_0^{(i)},c_1^{(i)},\cdots,c_N^{(i)})\quad (0\leq i \leq M),\\
A_{M+1}=
\begin{bmatrix}
e_0&w_{0,1}&\cdots&w_{0,N}\\
      &e_1      &\ddots&\vdots   \\
      &            &\ddots&w_{N-1,N}\\
      &            &          &e_N
\end{bmatrix},
\end{array}
\end{equation}
where $c_0^{(i)}=1$, $w_{m,n}=-\sum _{i=0}^M b_m^{(i)}c_n^{(i)}$,  
$({\rm tr} A_i=)\sum _{n=0}^N b_n ^{(i)} c_n^{(i)}=-\theta_i$ and $\sum _{i=0}^M b_n^{(i)} c_n^{(i)}=-\kappa_n$.
The essential number of the accessory parameters is confirmed to be $2MN$. 

The isomonodromic family of Fuchsian systems of the form (\ref{TsudaA})  is described by the integrability condition of the extended linear system, that is, (\ref{TsudaA}) itself and its deformation equations
\begin{equation}\label{TsudaB}
\frac{\partial \Phi }{\partial u_i} =B_i \Phi,\quad B_i=\frac{A_i}{u_i-z}-\frac{1}{u_i} \begin{bmatrix}
-\frac{\theta_i}{N+1}&         &                               \\
                               &\ddots&                               \\
                             * &         & -\frac{\theta_i}{N+1}
\end{bmatrix}\quad (1\leq i \leq M).
\end{equation}
In Theorem 7.2 in \cite{tsuda10}, it is proved that the integrability condition
\begin{equation}\label{ryoritsu:bibun}
\frac{\partial A}{\partial u_i} -\frac{\partial B_i}{\partial z}+[A,B_i]=0,
\end{equation}
of (\ref{TsudaA}) and (\ref{TsudaB}) is equivalent to the polynomial Hamiltonian system $\mathcal{H}_{N+1,M}$ via a certain change of variables.

The system in  the case $(M,N)=(M,1)$  coincides  with the Garnier system in $M$ variables, \cite{garnier12}, \cite{KO84} and thus in the case $(M,N)=(1,1)$ with the sixth Painlev\'{e} equation $P_{VI}$ \cite{Malm22}, \cite{Okamoto87}.
   It was  also shown that their solutions are expressed in terms of a generalization of hypergeometric functions $F_{N+1,M}$ given in (\ref{gauss:kakutyou}). 
   A linear Pfaffian system  of rank $MN+1$ which 
   $F_{N+1,M}$ 
   satisfies was derived. When $\kappa_0-\sum_{i=1}^M \theta =0$, the integrability condition (\ref{ryoritsu:bibun}) can be converted into the Pfaffian system for $F_{N+1,M}$. 
   In this way the hypergeometric solution for the system (see Thm. 3.2 in \cite{tsuda12} was obtained).

\end{document}